\documentclass[12pt]{article}
\newif\iffigs\figstrue
\usepackage{latexsym}
\iffigs
  \input{epsf}
\else
\fi
\font\tenmsbm=msbm10 scaled 1200
\font\sevenmsbm=msbm9
\newfam\msbmfam
\textfont\msbmfam=\tenmsbm
\scriptfont\msbmfam=\sevenmsbm
\def\msbm{\fam\msbmfam\tenmsbm}

\makeatletter
\@addtoreset{equation}{section}
\makeatother
\renewcommand{\theequation}{\thesection.\arabic{equation}}
\newcounter{parentequation}
\newenvironment{subequations}{%
  \refstepcounter{equation}%
  \begingroup 
  \let\protect\noexpand
  \edef\@tempa{\def\noexpand\theparentequation{\theequation}}%
  \expandafter
  \endgroup\@tempa
  \setcounter{parentequation}{\value{equation}}%
  \setcounter{equation}{0}%
  \def\theequation{\theparentequation\alph{equation}}%
  \ignorespaces
}{%
  \setcounter{equation}{\value{parentequation}}%
}
\newcommand{\eqn}[1]{(\ref{#1})}
\newsavebox{\uuunit}
\sbox{\uuunit}
    {\setlength{\unitlength}{0.825em}
     \begin{picture}(0.6,0.7)
        \thinlines
        \put(0,0){\line(1,0){0.5}}
        \put(0.15,0){\line(0,1){0.7}}
        \put(0.35,0){\line(0,1){0.8}}
       \multiput(0.3,0.8)(-0.04,-0.02){10}{\rule{0.5pt}{0.5pt}}
     \end {picture}}
\newcommand {\unity}{\mathord{\!\usebox{\uuunit}}}
\newsavebox{\bobox}
\sbox{\bobox}
    {\setlength{\unitlength}{0.825em}
     \begin{picture}(0.6,0.7)
        \thinlines
        \put(0,0){\line(1,0){0.7}}
        \put(0,0){\line(0,1){0.7}}
        \put(0.7,0){\line(0,1){0.7}}
        \put(0,0.7){\line(1,0){0.7}}
        \multiput(0,0)(0.05,0.05){14}{\rule{0.4pt}{0.4pt}}
        \multiput(0,0.65)(0.05,-0.05){13}{\rule{0.4pt}{0.4pt}}
     \end {picture}}
\newcommand {\boxtimes}{\mathord{\!\usebox{\bobox}}\,}

\def\IC{\hbox{\msbm C}}
\def\IZ{\hbox{\msbm Z}}

\def\bfone{\relax{\rm 1\kern-.35em 1}}
 \def\cB{{\cal B}}
\def\cC{{\cal C}} \def\cD{{\cal D}}
 
 \def\cG{{\cal G}}

\def\cN{{\cal N}}

\def\beq{\begin{equation}}
\def\eeq{\end{equation}}
\def\bea{\begin{eqnarray}}
\def\eea{\end{eqnarray}}
\def\bes{\begin{subequations}\begin{eqnarray}}
\def\ees{\end{eqnarray}\end{subequations}}
\def\bet{\begin{tabular}}
\def\eet{\end{tabular}}
\def\ol{\overline}
\def\nin{\noindent}
\newcommand{\dfrac}{\displaystyle \frac}
\def\a{\alpha}
\def\b{\beta}
\def\l{\lambda}

\def\g{\gamma}
\def\G{\Gamma}
\def\s{\sigma}
\def\e{\epsilon}

\def\bth{\bar{\theta}}
\def\hm{\hat{m}}
\def\hn{\hat{n}}

\begin{document}
\begin{titlepage}

\begin{flushright}
DFTT 99/67 \\
CERN--TH/99-388 \\
\end{flushright}

\vspace{2truecm}

\begin{center}

{ \Large \bf $M$--Theory on the Stiefel manifold and
3d Conformal Field Theories }

\vspace{1cm}

{ A. Ceresole$^{\S\star}$}\footnote{Permanent address: Dipartimento di Fisica,
Politecnico di Torino, C.so Duca degli Abruzzi, 24, I-10129 Torino. e-mail
:ceresole@polito.it, dauria@polito.it},
{G. Dall'Agata$^{\dag\star}$}\footnote{dallagat@to.infn.it},
{R. D'Auria$^{\S\star {1}}$}
and
{S. Ferrara$^{\S}$}\footnote{sergio.ferrara@cern.ch}

\vspace{1cm}

{$\S$ \it TH Division, CERN, 1211 Geneva 23, Switzerland \\
}

\vspace{1cm}

{ $\star$ \it
Istituto Nazionale di Fisica Nucleare, Sezione di Torino,\\
via P. Giuria 1, I-10125 Torino.
}

\medskip

{$\dag$ \it Dipartimento di Fisica Teorica, Universit\`a  di Torino,  \\
 via P. Giuria 1, I-10125 Torino.}

\medskip

\vspace{1cm}

\begin{abstract}
We compute the  mass and multiplet spectrum of $M$--theory compactified on the
product of $AdS_4$ spacetime by the Stiefel manifold $V_{(5,2)}=SO(5)/SO(3)$, and
we use this information to deduce via the AdS/CFT map the primary operator
content of the boundary  $\cN=2$ conformal field theory.
We make an attempt for a candidate supersymmetric gauge theory that, at strong
coupling, should be related to parallel $M2$--branes on the singular point of
the non--compact Calabi--Yau four--fold $\sum_{a=1}^5 z_a^2 = 0$, describing
the cone on $V_{(5,2)}$. 
\end{abstract}

\end{center}

\vskip 0.5truecm

Theories, Supergravity, Anti de Sitter Space,
Kaluza Klein Theories

\end{titlepage}
\newpage

\baselineskip 6 mm

\section{Introduction}

Duality between large N field theories on D3--branes placed at conical
singularities and Type IIB theory compactifications on a 5d manifold have
been the subject of many investigations in the last couple of years
\cite{JM}-\cite{WITT} (for a rewiew see \cite{rev}). 
In particular, in \cite{KW} an $\cN=1$ superconformal Yang--Mills field
theory has been constructed which turned out to be dual to Type IIB theory
compactified on the $T^{11}=SU(2)^2/U(1)$ manifold. 
Solid  evidence on such duality was developed in \cite{KW2,G,JR} and
in \cite{CDDF}, where a one--to--one correspondence was established between the
Kaluza--Klein (KK) supergravity spectrum on $T^{11}$ \cite{CDD}  and the
boundary superconformal operators.  In the case of $M2$ and $M5$ branes the
$AdS/CFT$ conjecture was first studied for maximal supersymmetry in
compactifications on spheres $S^7 $ and $S^4 $ \cite{AOY,halyo, Min} and then
further extended to lower $\cN$ for sphere-orbifolds and other brane  systems
\cite{orb,IMSY}.

In \cite{noi} the question was raised whether one could similarly construct a
SuperConformal Field Theory (SCFT) in three dimensions which is dual for
large brane-flux $N$ to $M$--theory compactified on $AdS_4\times M_7$
\cite{noi,fig,MP}, where $M_7$ is one among the compact Einstein manifolds that
were classified \cite{CRW} and analysed in the eighties \cite{Duff, book}.
Such three--dimensional SCFT's would then be defined as the conformal limit of
the world--volume theory of N M2--branes positioned at a conical singularity of
$M_3\times Y_8$, where $Y_8$ is the cone over $M_7$.

Three-dimensional SCFT's are more difficult to analyze because they
emerge in non-perturbative limit of conventional gauge quantum field
theories.
The origin of this difficulty is well known, and can be traced back to the
fact that the three dimensional bare gauge coupling constant is dimensionful,
so that a conformal description of the theory dual to supergravity is only
possible in the infrared limit, where the gauge coupling blows up and the
gauge fields are integrated out. 

In this regime, the SCFT is described as the low energy theory of N
coincident M2--branes, where N is the number of units of flux of the dual of
the eleven--dimensional four--form on the internal manifold $M_7$. 
Indeed, compactifications of $M$--theory on a circle leads to a D2--brane
description in type IIA theory whose world--volume gauge theory is not
conformal and whose near horizon geometry generically is not $AdS$
\cite{IMSY,PS,rev}. The superconformal description is recovered in the strong
coupling limit of type IIA decompactification, when $g^2_{eff}>>N^{1/2}$,
$u<<g^2_{YM}$, where $g^2_{eff}=g_s N/u$ \cite{rev} and $u$ is the energy
scale.  
When the radius of the circle goes to infinity, the gauge coupling also blows
up and we reach the M2--brane description in the infrared of the
three--dimensional gauge theory. Its relevant degrees of freedom are given in
terms of the d=3 chiral multiplets and are related to the KK excitations of
$d=11$ supergravity.
Vice--versa, if $g^2_{eff}\rightarrow 0$ and $u\rightarrow \infty$, we are
describing the ultraviolet limit of the d=3 Yang--Mills theory (for a thorough
discussion of the above considerations see \cite{IMSY,PS,rev}).

It is also important to observe that, although the $d=3$ ultraviolet
gauge theory has both a Coulomb and a Higgs branch, in the  $AdS_4/CFT_3$
correspondence one is mainly interested in the Higgs branch, parametrized by
the {\sl vev}'s of the  scalars of the chiral multiplets. 
The Coulomb branch scalars, belonging to the vector multiplets, are excluded
since their {\sl vev}'s can be safely put to zero.

In \cite{Fetc} by a thorough study of the KK excitations and $OSp(4|2)$
multiplets \cite{CFN} on the $M^{111}=SU(3)\times SU(2)\times
U(1)/SU(2)\times U(1)\times U(1)$ \cite{M111} and $Q^{111}=SU(2)^3/U(1)^2$
\cite{Q111} manifolds, the relevant $\cN=2$ SCFT's  have been constructed on
the basis of the mass spectrum, and conjectured to be dual to $M$-theory
compactifications.

In the above construction, it was important that both $M^{111}$ and $Q^{111}$
admit a description in terms of toric geometry \cite{D}.  
This allowed to identify the fundamental degrees of freedom of the underlying
SCFT and thus to find the abelian gauge theories, whose moduli space of vacua
(the Higgs branch component) is isomorphic to the conifolds over the two
seven--manifolds.

This paper  analyses  the case of M--theory compactified on the real Stiefel
manifold $V_{(5,2)}\equiv SO(5)/SO(3)$ with $SO(3)$ canonically embedded in
$SO(5)$. The relevant four-fold cone whose base is $V_{(5,2)}$ was identified
in \cite{KW}.

The Stiefel manifold is  peculiar among the Einstein spaces leading to
$\cN=2$ supersymmetry in that it does not admit a toric description \cite{D}.
Nevertheless, we shall see that it is possible to build up the {\sl single
brane} ($N=1$) theory and to conjecture its $N>1$ extension by performing the
following steps. 

We first analyse the full KK mass spectrum and  $OSp(2|4)$ supermultiplets
for the $V_{(5,2)}$ manifold: this is in complete analogy with the procedure
in \cite{CDDF} for $T^{11}$ and in \cite{Fetc,M111} for some M-theory examples.
A relevant point of our analysis with respect to \cite{Fetc} is the absence
of Betti multiplets since there are no non--trivial Betti numbers on
$V_{(5,2)}$ 
\beq 
b_p=0,\qquad\qquad p=1,\ldots,6,\qquad\qquad b_0=b_7=1.
\eeq
Thus, there is no continuous baryonic symmetry \cite{Wittbaryon,GK,MP,CDDF,CDD}
in the corresponding SCFT. 

The second step is to construct admissible superconformal boundary
operators to be put in dual correspondence with those KK
supermultiplets undergoing shortening when the required unitarity
bounds on the $OSp(4|2)$ representations are saturated.
To arrive to a well founded conjecture for the SCFT operators, we are guided
by the consideration of the classical equation describing the eight
dimensional cone $Y_8$ over $V_{(5,2)}$. 
It turns out that the solution of the cone equation \cite{KW}
\beq
\sum_{a=1}^4 z_a^2=0
\eeq
can be obtained in the simplest way by use of the so
called Pl\"ucker embedding, where  beside the pfaffian identity on the
coordinates $p_{ij}$ = $-p_{ji}$ (where $i,j$ are indices in the fundamental
representation of $Sp(4)\equiv SO(5)$)
\beq
\label{pfaf}
\epsilon^{ijkl} p_{ij} p_{kl}=0
\eeq
one also uses the traceless constraint
\beq
\label{trac}
C^{ij}p_{ij}=0.
\eeq
As shown in \cite{Fetc}, the solution to \eqn{pfaf} can be given as 
\beq
p_{ij}=A_{[i} B_{j]}.
\eeq
In the present case, we find that one can write the coordinates parametrizing
the cone $Y_8$ as the bilinear
\beq
\label{za}
z^a=A^i\ \Gamma^a_{ij}  B^j,
\eeq
where $A^i$, $B^i$ are in the $\bf 4$ of $Sp(4)\equiv SO(5)$ and
$\Gamma^a_{ij}$ are the gamma matrices in five dimensions.
The vanishing of a $SU(2)$ D--term  fixes the residual $SL(2,\IC)$ invariance
of equations \eqn{pfaf} and \eqn{trac}. 
The above solution \eqn{za} is obtained by a procedure quite analogous  to
that employed in \cite{KW} to solve the cone equation on $T^{11}$ in terms of
two objects $A^i$, $B^i$ belonging to the representation $(1/2,0)\oplus
(0,1/2)$ of $SU(2)\times SU(2)$.  
There, the analogous of $SL(2,\IC)$ invariance was given by equation (13) of
\cite{KW}, namely invariance under complex rescalings.

This discussion gives us a little but useful information on the
gauge group $\cG$ of the theory in the ultraviolet regime.
More precisely, since the equation for the vanishing of the $D$--terms is
$SU(2)$ valued, the gauge group should reduce to $SU(2)$ for a single
$M2$--brane at the conical singularity. 
Then, in virtue of the fact that the conifold coordinate $z^a$  does indeed
appear in the KK spectrum and is a gauge singlet, we arrive at the conclusion
that for $N>1$ the basic singleton $S^\alpha_i$, transforming in the $\bf 4$
of $SO(5)$, must be in a pseudoreal representation of $\cal G$ labeled by the
index $\alpha$. 
Albeit this requirements could be satisfied by several groups,
we are led to conjecture the simple choice ${\cal G}=USp(2N)\times O(2N-1)$,
where the index $\alpha$ of $S^\alpha_i$ belongs to the bifundamental
representation\footnote{Indeed such gauge group appear in the
context of orientifold models \cite{uranga}.} of $\cal G$.

Once we have some solid base for the choice of the basic degrees of freedom in
the dual three--dimensional $\cN=2$ SCFT, we can perform the last step and
show that it is possible to construct a complete set of conformal primary
operators {\sl matching all} the KK multiplets previously obtained.

In establishing such correspondence we follow the procedure already
employed in \cite{CDDF} for the $T^{11}$ case. In particular we find that,
as for $T^{11}$, there are long multiplets with rational protected
dimensions\footnote{ The same feature was also found in \cite{Fetc}.}.

It is interesting to note that since in the infrared limit the gauge
field is integrated out, one may expect that it should be related in the SCFT
to some composite field in terms of the singleton $S^\alpha_i$.
In fact  we find in the gravitino sector a composite
field $X_\alpha$
obeying $\bar D^\alpha X_\alpha=0$ whose $\bar\theta$ component\footnote{
For three--dimensional superfields  we define:
$\theta_\a=\theta_\a^1+i\theta_\a^2$,
$\bar\theta_\a=\theta_\a^1-i\theta_\a^2$.
Conformal dimensions $\Delta$ and $R$--symmetry $y$ quantum numbers are
$(\Delta=1/2,y=1)$ for $\theta_\a$ and $(\Delta=1/2,y=-1)$ for
$\bar\theta_\a$.} has the right quantum numbers of a gauge field, being a
singlet of the flavour group $SO(5)$ and having $R$--symmetry $y=0$.

In the rest of the paper, section 2 briefly deals with the harmonic analysis on
the Stiefel manifold while section 3 contains the results on the full mass
spectrum and its assembling into $OSp(4|2)$ supermultiplets, with particular
emphasis on the shortening patterns due to saturation of unitarity bounds.

Section 4, relying on the solution of the conifold equation, proposes a
candidate for the classical $N=1$ theory which is supported by the
condition of vanishing D--terms. The $N>1$ extension is then
conjectured to be related in the ultraviolet to a gauge theory
of $D2$-branes given by the product of two non--simply laced groups
$\cG=USp(2N)\times O(2N-1)$ with chiral multiplet $S_i^{\alpha}$ transforming
in the spinor representation of the flavour group $SO(5)$ and in the
bifundamental of $\cG$.

Some considerations are also given on the possible existence of
a superpotential, at least in the $N=1$ case. In section 5, after a short
summary of the conformal operators related to the shortenings of
KK representations, we construct a set of conformal operators which can be
put in correspondence with the various supermultiplets.
Finally, we give a summary of our result in Section 5 while
some of the more technical aspects regarding useful tools for the harmonic
analysis are contained
in two appendices.

\section{The mass spectrum of $V_{(5,2)}$}

Harmonic  analysis on the coset space  $V_{(5,2)}\equiv SO(5)/SO(3)$  yields
the complete mass spectrum, which as in the other $\cN=2$ supersymmetric
compactifications, is arranged into $Osp(4|2)$ representations.
Referring to \cite{CDD} and references therein for the relevant details
concerning harmonic expansion on a generic coset  manifold, we give below
the essential ingredients for carrying out the computations and collect our
results in tables for the various supermultiplets.

As in any KK compactification, after the linearization of the
equations of motion of the field fluctuations, one is left with a
differential equation on the eleven--dimensional fields
$\Phi_{[\l_1,\l_2,\l_3]}^{\{J\}} (x,y)$ 
\beq
\label{LB}
(\Box^{\{J\}}_x + \boxtimes^{[\l_1,\l_2,\l_3]}_y)
\Phi^{\{J\}}_{[\l_1,\l_2,\l_3]}(x,y) = 0.
\eeq
Here the field $\Phi_{[\l_1,\l_2,\l_3]}^{\{J\}} (x,y)$ depends on the
coordinates $x$ of $AdS_4$ and $y$ of $V_{(5,2)}$, and transforms
irreducibly in the representations $\{J\}$ of $SO(3,2)$ and
$[\l_1,\l_2,\l_3]$ of $SO(7)$. $\Box_x$ is the kinetic operator for a field of
quantum numbers $\{J\}\equiv \{\Delta,s\}$ in four--dimensional $AdS$ space
and $\boxtimes_y$ is the Laplace--Beltrami operator for a field of spin
$[\l_1,\l_2,\l_3]$  in the internal space $V_{(5,2)}$.
(In the following we mostly omit the index $\{J\}$ on the fields.
The symbol $[\l_1,\l_2,\l_3]$ denotes the quantum numbers of the $SO(7)$
representation in the Young tableaux formalism.)

More specifically, one expands the $d=11$ supergravity fields
$\Phi_{[\l_1,\l_2,\l_3]} (x,y)$=$\{h_{\hat a \hat b}$, $A_{\hat a\hat b
\hat c}$,$\psi_{\hat a}\}$ ($\hat a =(a,\alpha)$, $a=1,\ldots,4$,
$\alpha=1,\ldots,7$) in the harmonics of $V_{(5,2)}$ transforming irreducibly
under the isometry group of $V_{(5,2)}$,  and computes the action of
$\boxtimes_y$ on these harmonics. The eigenvalues are simply related
to the $AdS$ mass.

The two necessary ingredients in this computation are the geometric structure
and the harmonics of the coset space.

\medskip

\nin
{\it Geometry}

\medskip

\nin
We give  here a brief description of the essential geometrical elements
of the Stiefel manifold such as the metric and the Riemmanian curvature,
that  are used to build the invariant Laplace--Beltrami operators.

We remind that the Stiefel manifold, beside $SO(5)$, has an extra isometry
$SO(2)_R$ that can be identifyed with the $R$--symmetry group \cite{CRW}
\beq
\frac{G}{H}= V_{(5,2)} \equiv \frac{SO(5) \times SO(2)_R}{SO(3) \times
SO(2)_H},
\eeq
where the embedding of the $SO(2)_{H}$ into $SO(5) \times SO(2)_R$ is
diagonal and the embedding of $SO(3)$ in $SO(5)$ is the canonical one,
namely the fundamental of $SO(5)$ breaks under $SO(3)$ as ${\bf 5}\rightarrow
{\bf 3}+{\bf 1}+{\bf 1}$ (other embeddings yield different inequivalent $M_7$
with $\cN\neq 2$).

We call $\Lambda, \Sigma = 1, \ldots 5$ the $SO(5)$ indices, $I =
1,2,3$ the $SO(3)$ indices and $A = 1,2$ the $SO(2)$ ones.  The
adjoint generators of $SO(5)$ and $SO(2)_R \simeq U(1)_R$ are
respectively $T^{\Lambda \Sigma}, U$.

The $SO(5)$ algebra is
\beq
[T_{\Lambda\Sigma}, T_{\Gamma\Delta}]  = - \eta_{\Lambda \Gamma}
T_{\Sigma \Delta} +
\eta_{\Lambda \Delta} T_{\Sigma \Gamma} - \eta_{\Sigma \Delta}
T_{\Lambda \Gamma}
+\eta_{\Sigma \Gamma} T_{\Lambda \Delta},
\eeq
which means that our generators in the vector representation are
$(T^{\Lambda\Sigma})_{\Gamma\Delta} = 2\delta^{\Lambda\Sigma}_{\Gamma\Delta}$.

For the canonical embedding of $SO(3)$ in $SO(5)$, $\Lambda = (I,A)$, we can
define $J_K \equiv \frac{1}{2} \e_{KIJ} T_{IJ}$ as the $SO(3)$ generators and
$ N \equiv T_{45} + U$ as the $SO(2)_{H}$ generator. 
The coset generators are given by $T_{IA} \equiv (T_{m},T_{\hm})$ ($m,\hm =
1,2,3$), $T_7 \equiv T_{45} - U$.

Since the vielbeins are coset--algebra valued, we use the same convention
for labelling the vielbein directions in the coset space.

Given the structure constants ${C^a}_{bc}$ of the coset, the Riemann tensor
is defined by the formula 
\bea
\label{Riemannbis}
{R^a}_{bde} &=& - \frac{1}{4} {\IC^a}_{bc} {C^c}_{de} \frac{r(d)
r(e)}{r(c)} - \frac{1}{2}
{C^a}_{bi}{C^i}_{de} r(d)r(e) + \nonumber \\
&-& \frac{1}{8} {\IC^a}_{cd} {\IC^c}_{be}
+ \frac{1}{8} {\IC^a}_{ce} {\IC^c}_{bd},
\eea
whose derivation we give in  Appendix A.
Here we simply point out that the $r(a)$ are the rescalings of the
vielbeins needed to obtain an Einstein space and that the
${\IC^a}_{bc}$ are certain specific combinations of the structure
constants.

We have imposed\footnote{ Note that there is an ambiguity in the sign of the
rescalings, since the Einstein space requirement on the
curvature determines only their square.  However, this ambiguity is
only apparent.  
While the partially reflected solutions with $r(a) \to - r(a)$ or $r(b) \to -
r(b)$ are perfectly equivalent to our description, a change in the sign of
$r(c)$ implies  that we really reflect the orientation on the manifold and as
a consequence we completely break supersymmetry.}
\beq
\label{rescalings}
r(a) = r(b) = 4\sqrt{\frac{2}{3}} \, e, \qquad r(c) = -\frac{4}{3} \, e
\eeq
to obtain ${R^{a}}_{b} = 12 e^2 \, \delta^{a}_{b}$.
With such rescalings \eqn{rescalings}, we obtain
\bea
R^{mn}{}_{kl} &=& R^{\hm\hn}{}_{\hat{k}\hat{l}} = \frac{32}{3}\, \delta^{mn}_{kl} \\
R^{mn}{}_{\hat{k}\hat{l}} &=& \frac{20}{3} \delta^{mn}_{kl} \\
R^{m\hn}{}_{k\hat{l}} &=& \frac{4}{3} \delta^{mn} \delta_{kl} - 2 \delta^m_l
\delta^n_k\\
R^{7m}{}_{7n} &=& R^{7\hm}{}_{7\hn} = 2 \delta^{m}_{n}.
\eea

\medskip

\nin
{\it The harmonics}

\medskip

\nin
The harmonics on the coset space $V_{(5,2)}$ are labelled by two kind
of indices, the first giving the specific representation of the
isometry group $SO(5) \times U_{R}(1)$ and the other referring to the
representation of the subgroup $H \equiv SO(3) \times SO(2)$.
The harmonic is thus denoted by $Y^{(M,N,Q)}_{(q_H)}(y)$, where $M$,$N$
are the quantum numbers of the $SO(5)$ representation, $Q$ is the
$U_{R}(1)$ charge and $q_H$ are the $H$--quantum numbers.

The above results imply that an $SO(7)$ field $\Phi_{[\l_1,\l_2,\l_3]}(x,y)$
can be splitted into the direct sum of $H$ irreducible fragments labelled by
$q_H$. The analysis of the reduction of the $SO(7)$ representation
under the $H$ group reported in Appendix B, yields that the
vector and spinor $SO(7)$ representations break as
\beq
\label{decco}
\begin{array}{rcll}
{\bf 7} &\to& {\bf 3}_1 \oplus {\bf 3}_{-1} \oplus {\bf 1}_0, \\
{\bf 8} &\to& {\bf 3}_{1/2} \oplus {\bf 3}_{-1/2} \oplus {\bf 1}_{3/2}
\oplus {\bf 1}_{-3/2},
\end{array}
\eeq
and by taking suitable combinations one can also derive all the other tensor
decompositions.

The generic field $\Phi_{[\l_1,\l_2,\l_3]}(x,y)$  can then be expanded as
follows 
\beq
\label{hexp}
\Phi_{ab\ldots}(x,y) = \sum_{(\nu)} \sum_{(m)} \Phi_{(\nu)(m)}(x)
Y^{(\nu)(m)}_{ab\ldots} (y),
\eeq
where $a,b,\ldots$ are $SO(7)$ tensor (or spinor) indices of the
representation $[\l_1,\l_2,\l_3]$, $(\nu)$ is a shorthand notation
for $(M,N,Q)$ and $m$ labels the representation space of ($M,N,Q$).

Of course, not all the harmonics are allowed in the \eqn{hexp}
expansion, as the irrepses of $SO(7)$ appearing in \eqn{hexp} must contain,
once reduced with respect to $H$, at least one of the representations
appearing in the decomposition of $[\l_1,\l_2,\l_{3}]$ under  $H$. 
This gives some constraints on $M,N,Q$ which select the allowed
representations $(\nu)$.

We write a generic representation of $SO(5)$ in the Young
tableaux formalism $[\l_1,\l_2] = [M+N,M]$:
\bea
\label{Tab1}
& &
\begin{array}{l} \begin{array}{|c|c|c|c|c|c|c|c|c|}\hline  &  &
$\ldots$ &  & & & $\ldots$ &
\\\hline \end{array} \nonumber \\
 \begin{array}{|c|c|c|c|}  & & $\ldots$ & \\\hline
 \end{array}
 \end{array} \\
& & \;\, \underbrace{\hspace{1.9cm}}_{M}
\underbrace{\hspace{1.9cm}}_{N}
\nonumber
\eea
and $Q$, the $U_{R}(1)$ charge, is defined by the $U_R(1)$ harmonic
$e^{i Q\,  \phi}$.

A specific component of \eqn{Tab1} can then be written as
\bea
&&
\begin{array}{l}
\begin{array}{|c|c|c|c|c|c|c|c|c|}\hline
  4  & 4  & $\ldots$ & \,i\, & \,j\,& \hbox{{\footnotesize +}}
  & $\ldots$ & -
 \\\hline \end{array} \nonumber \\
 \begin{array}{|c|c|c|c|} 5 & 5 & $\ldots$ & \hbox{{\footnotesize +}} \\\hline
 \end{array}
\end{array}
\nonumber
\eea
where we defined the $U_R(1)$ fixed charge combinations
$$
\bet{|c|}\hline {\footnotesize $\pm$} \\\hline \eet \equiv
\bet{|c|}\hline 4
\\\hline\eet \pm i \bet{|c|}\hline 5
\\\hline\eet
$$

\bigskip

\begin{center}
\begin{tabular}{|c|c|c|c|c|c|}\hline
spin&$\Phi (x)$          & $\Phi (x,y)$    & harmonic&
$\boxtimes$ operator $\phantom{\stackrel{V}{V}}$ &$SO(7)$ irrep\\  \hline\hline
2   & $h_{ab}$         & $h_{ab}$     & Y
&$\Box =\cD^\a \cD_\a$& [0,0,0]  \\
\hline
$1^+$ & $A_a$, $W_a$    & $h_{a\beta}$, $A_{ab\gamma}$&$Y_\alpha$& $\Box +24$
&[1,0,0]\\
$1^-$ & $Z_a$           & $A_{a\beta\gamma}$&$Y_{[\beta\gamma]}$ &
$(\Box +40)\delta_{\beta\gamma}^{\mu\nu}-2 \cC^{\mu\nu\beta\gamma}$
&[1,1,0] \\\hline
$0^+$ & S,$\Sigma$        & $h_{ab}$, $h_{\alpha\beta}$,
$A_{abc}$ & Y & $\Box$&[0,0,0] \\
    & $\phi$            & $h_{\alpha\beta}$ & $Y_{(\alpha\beta)}$ &
$(\Box +40)\delta_{(\beta\gamma)}^{(\mu\nu)}-4 \cC^{\mu\beta\nu\gamma}$
&[2,0,0] \\
$0^-$ & $\pi$             & $A_{\alpha\beta\gamma}$ &
$Y_{[\alpha\beta\gamma]}$
&$1/24 \epsilon^{\mu\nu\rho\s\alpha\beta\gamma}\cD_\s$
&[1,1,1] \\
\hline\hline
3/2 & $\chi_a$        &$\psi_a$ & $\Xi$ &$\cD\!\!\!\slash -7 \phantom{\stackrel{V}{V}}$ &[
$\frac{1}{2},\frac{1}{2},\frac{1}{2}$]
\\
1/2 &$\lambda_L$      &$\psi_a$,$\psi_\alpha$&$\Xi$, $\Xi_\alpha$ & &\\
    &$\lambda_T$      &$\psi_\alpha$ & $\Xi_\alpha $ &
$\cD\!\!\!\slash -5 \phantom{\stackrel{V}{V}}$ &[$   \frac{3}{2},\frac{1}{2},\frac{1}{2}$]\\
\hline
\end{tabular}

\vskip .3cm

{\bf Table 1: Correspondence between 11d and 4d fields
  and harmonics \cite{DF}}
\end{center}

One can now proceed with the KK analysis, implementing the
above informations in all the Laplace--Beltrami
operators and computing the eigenvalues of the various harmonics
and thus the masses of the KK states.

Although straightforward in principle, this analysis can become quite
cumbersome for some of the higher spin operators. 
Luckily, it is not really necessary to complete the whole task. 
In fact, due to the $\cN=2$ supersymmetry,   this peculiar compactification
falls in the class considered in \cite{CFN}, where the $OSp(4|2)$ multiplet
structure was  elucidated, together with the mass values expected for states
of given quantum numbers. 
Most of these multiplets can be filled by using only our results for the
simpler operators, while the entries in the remaining slots can be determined
with the help of supersymmetry.

From the scalar, spinor and vector eigenvalues, we have obtained the
masses for all the graviton and gravitinos and for some of the vectors, spinors
and scalars, which let us fill  the five types of supermultiplets presented in
Tables 2--6, with all  the shortening patterns.
A preliminary analysis of the rank of the two--form matrix yields
that we can have at most two more vector multiplets,
which we guess do not undergo shortening.

Indeed, as in the $T^{11}$ case \cite{CDD,CDDF},
all the mass eigenvalues depend on  the $G$--quantum numbers only through the
function $H_{0}$, which is the scalar laplacian eigenvalue.
For the Stiefel manifold, such eigenvalue is given by
\beq
H_{0}(M,N,Q) = \frac{32}{9} \left( 6 M^2 + 9 N + 3 N^2 + 12 M + 6MN -
Q^2\right).
\eeq
Since for a given  number of preserved supersymmetries,
the structure of the linearized equations, supersymmetry relations and
supermultiplets are fixed, we can  suppose that also the mass formulae
in terms of $H_0$ are universal for all seven-dimensional $\cN=2$
supersymmetric compactifications.
By this we mean that not only the number and type of multiplets for
different compactifications are the same, but also the $H_0$ mass
dependence  should be equal.

This is exactly what we find by comparing our  case with
the $M^{111}$ compactification \cite{M111}, and we
expect such agreement to hold also for the $Q^{111}$ space.
Of course, the shortening patterns as well as the possible presence
of Betti multiplets will be model dependent features,
as they derive either from certain functions of $H_0$ taking rational
values or from non--trivial Betti numbers of the relevant manifold.
However, in this respect, the two vector multiplets mentioned before
have $\Delta = \frac{1}{2} +\frac{1}{4}\sqrt{4+H_0}$ and $\Delta = \frac{1}{2}
+\frac{1}{4}\sqrt{H_0(M,N,Q+\frac{3}{2})-28}$, and satisfy no shortening
conditions. This makes us confident that {\it all} the relevant output
derived from the supergravity analysis is correct.

\section{The $AdS_4 \times V_{(5,2)}$ multiplet structure.}

We report below in tables 2--6  the five families of supermultiplets
we have found:
one graviton multiplet, two gravitino multiplets and two
vector multiplets.

Each table has five main columns.
The first column contains the spin quantum number of the state, while
in the second  we give its  $\Delta^{(s)}$ value.
The basic value of $\Delta$ assigned to each multiplet is the one
belonging to a vector field, a spin 1/2  or  a scalar for
the graviton, gravitino and vector multiplets respectively.
In the third column we write the $R$--symmetry of the state where the value
$y$ is assigned to the state with $\Delta^{(s)}=\Delta$ .
We use here $y = \frac{2}{3} Q$, since this varies in integer steps according
with the usual convention on the unit value of the $R$--charge of the $\theta$
coordinate \cite{CFN}.
The fourth column shows the specific field of the KK spectrum that is
associated with the given $OSp(4|2)$ state, according
to the notations of \cite{CFN}.
The fifth column contains the mass of the state\footnote{ We give
the value of the mass for the fermions and the mass squared for the bosons.}
given in terms of $H_0$.

For generic $SO(5)$ quantum numbers and  $R$--symmetry values,
the multiplets of tables 2--6 are long multiplets of $OSp(4|2)$.
However, group theory predicts \cite{CFN} shortening
in correspondence with specific threshold values
of the  quantum numbers. These give rise to chiral ($\bullet$),
semi--long ($\star$) or massless ($\diamond$) multiplets. The above symbols
appear in the extra left columns to denote the surviving states in the
shortened multiplets. In absence of these symbols no shortening
of any kind can occur for that multiplet.

\begin{center}

 \begin{tabular}{|c|c|c|c|c|c|c|c|}\hline
  & & spin &   $\Delta^{(s)}$     &   $R$--symm.   &   field & Mass      \\
 \hline
 \hline
 $\diamond$ &$\star$ & 2     & $\Delta+1$      &$y$      &$h$
     & $H_0$               \\\hline
$\diamond$ & $\star$ &  3/2    &$\Delta+1/2$     &$y-1$    &$\chi^+$
&$-6+\sqrt{H_0+36}$     \\
$\diamond$ & $\star$ &3/2          &$\Delta+1/2$     &$y+1$
&$\chi^+$    &          $-6+\sqrt{H_0+36}$     \\
 &  $\star$ &3/2          &$\Delta+3/2$     &$y-1$    &$\chi^-$   &
     $-6-\sqrt{H_0+36}$   \\
 &   & 3/2      &$\Delta+3/2$     &$y+1$    &$\chi^-$
&$-6-\sqrt{H_0+36}$      \\
 \hline
 $\diamond$ & $\star$ & 1     &$\Delta$  &$y$
&$ A/W$ &$H_0+48-8\sqrt{H_0+36}$    \\
  &  & 1     &$\Delta+1$&$y+2$  &$Z$  &$H_0+32$  \\
  & $\star$ &1    &$\Delta+1$&$y-2$  &$ Z$   & $H_0+32$  \\
  & $\star$ & 1     &$\Delta+1$&$y$    &$Z$   &$H_0+32$  \\
  &      & 1     &$\Delta+1$&$y$    &$Z$   &$H_0+32$   \\
  &    & 1     &$\Delta+2$&$y$    &$A/W$   &$H_0+48+8\sqrt{H_0+36}$   \\
\hline
  &  &1/2 & $\Delta+1/2$      &$y+1$  & $\lambda_T^-$&$2-\sqrt{H_0+36}$  \\
  & $\star$ &1/2& $\Delta+1/2$      &$y-1$  & $\lambda_T^-$&$2-\sqrt{H_0+36}$
  \\
& &1/2  & $\Delta+3/2$     &$y-1$  & $\lambda_T^+$&$2+\sqrt{H_0+36}$
\\
& &1/2  & $\Delta+3/2$ &$y+1$  &$\lambda_T^+$  &$2+\sqrt{H_0+36}$  \\
\hline
 &   &0  & $\Delta+1$&$y$ & $ \phi$ & $H_0+32$     \\
\hline
 \end{tabular}
\vskip .3cm
{\bf Table 2: Long Graviton Multiplet} \qquad $\Delta=\frac{1}{2}+
\frac{1}{4}\sqrt{H_0+36}$.

\bigskip

 \begin{tabular}{|c|c|c|c|c|c|c|}\hline
  & spin &   $\Delta^{(s)}$     &   $R$--symm.   &   field & Mass      \\
 \hline \hline
$\star$ & 3/2 &$\Delta+1$&$y$&$\chi^+$&$-8+\sqrt{H_0+24}$\\
 \hline
   $\star$ &1        &$\Delta+1/2$     &$y+1$    &$A/W$   &
$H_0+56 - 12\sqrt{H_0+24}$   \\
   $\star$ &1        &$\Delta+1/2$     &$y-1$    &$A/W$   &
$H_0+56 - 12\sqrt{H_0+24}$   \\
   $\star$ &1 &$\Delta+3/2$     &$y-1$    &$ Z$
&$H_0+24 - 4\sqrt{H_0+24}$  \\
    &1 &$\Delta+3/2$     &$y+1$    &$ Z$
&$H_0+24- 4\sqrt{H_0+24}$     \\
\hline
  $\star$ &1/2    &$\Delta$  &$y$    &$\lambda_T^+$  &$-8+\sqrt{H_0+24}$    \\
  $\star$ &1/2 &$\Delta+1$&$y-2$  &$\lambda_T^-$   &$4-\sqrt{H_0+24}$  \\
  $\star$ &1/2 &$\Delta+1$&$y$  & $ \lambda_T^-$   & $4- \sqrt{H_0+24} $  \\
    &1/2 &$\Delta+1$&$y+2$    &$\lambda_T^-$    &$4-\sqrt{H_0+24}$   \\
    &1/2 &$\Delta+1$      &$y$  &$\lambda_T^-$  &$4-\sqrt{H_0+24}$    \\
   &1/2 &$\Delta+2$      &$y$  &$\lambda_T^+$  &$\sqrt{H_0+24}$    \\
 \hline
 $\star$ &0  & $\Delta+1/2$      &$y-1$  &$\pi$ & $H_0+56- 12\sqrt{H_0+24}$  \\
 &0 &$\Delta+1/2$      &$y+1$  &$\pi$  &$H_0+56-12\sqrt{H_0+24}$ \\
    &0 & $\Delta+3/2$      &$y+1$  & $\phi$&   $H_0+24- 4\sqrt{H_0+24}$   \\
  &0 &$\Delta+3/2$      &$y-1$  &$\phi$  &  $H_0+24- 4\sqrt{H_0+24}$  \\
\hline
 \end{tabular}

\vskip .3cm

{\bf Table 3: Long Gravitino Multiplet I}\qquad
$\Delta=-\frac{1}{2}+\frac{1}{4} \sqrt{H_0 + 24}$

\bigskip

 \begin{tabular}{|c|c|c|c|c|c|c|}\hline
  & spin &   $\Delta^{(s)}$     &   $R$--symm.   &   field & Mass      \\
 \hline \hline
$\star$ & 3/2 &$\Delta+1$&$y$&$\chi^-$&$-8-\sqrt{H_0+24}$\\
 \hline
   $\star$ &1        &$\Delta+1/2$     &$y+1$    & $Z$
&$H_0+24+ 4\sqrt{H_0+24}$  \\
   $\star$ &1        &$\Delta+1/2$     &$y-1$    &$Z$
&$H_0+24+ 4\sqrt{H_0+24}$   \\
   $\star$ &1 &$\Delta+3/2$     &$y-1$    &$A/W$
&$H_0+56+ 12\sqrt{H_0+24}$     \\
   &1 &$\Delta+3/2$     &$y-1$    &$A/W$         &
$H_0+56+ 12\sqrt{H_0+24}$  \\
\hline
  $\star$ &1/2    &$\Delta$  &$y$    &$\lambda_T^-$  &$-\sqrt{H_0+24}$    \\
  $\star$ &1/2 &$\Delta+1$&$y-2$  &$\lambda_T^+$   &$4+\sqrt{H_0+24}$  \\
  $\star$ &1/2 &$\Delta+1$&$y$  & $ \lambda_T^+$   & $4+ \sqrt{H_0+24} $  \\
    &1/2 &$\Delta+1$&$y+2$    &$\lambda_T^+$    &$4+\sqrt{H_0+24}$   \\
    &1/2 &$\Delta+1$      &$y$  &$\lambda_T^+$  &$4+\sqrt{H_0+24}$    \\
   &1/2 &$\Delta+2$      &$y$  &$\lambda_T^-$  &$-8-\sqrt{H_0+24}$    \\
 \hline
 $\star$ &0    & $\Delta+1/2$      &$y-1$  &$\phi$ &$H_0+24+
4\sqrt{H_0+24}$   \\
 &0 &$\Delta+1/2$      &$y+1$  &$\phi$  &$H_0+24+ 4\sqrt{H_0+24}$      \\
    &0 & $\Delta+3/2$      &$y+1$  & $\pi$&$H_0+56+ 12\sqrt{H_0+24}$   \\
  &0 &$\Delta+3/2$      &$y-1$  &$\pi$  &$H_0+56+ 12\sqrt{H_0+24}$    \\
\hline
 \end{tabular}

\vskip .3cm

{\bf Table 4: Long Gravitino Multiplet II}\qquad
$\Delta=\frac{3}{2}+\frac{1}{4} \sqrt{H_0 + 24}$

\bigskip

 \begin{tabular}{|c|c|c|c|c|c|}\hline
 spin & $\Delta^{(s)}$ &   $R$--symm.   &   field & Mass      \\
 \hline
 \hline
 1   &$\Delta+1$      &$y$
&$A/W$&$H_0+96 + 16\sqrt{H_0+36}$\\
 \hline
1/2          &$\Delta+1/2$     &$y-1$    &$\lambda_{T}$
&$6 + \sqrt{H_0+36}$  \\
1/2               &$\Delta+1/2$     & $y+1$    & $\lambda_{T}$          &
$6 +  \sqrt{H_0+36}$     \\
1/2          &$\Delta+3/2$     &$y-1$    &$\lambda_{L}$
 &$10 + \sqrt{H_0+36}$  \\
1/2&$\Delta+3/2$     &$y+1$    &$\lambda_{L}$   &$10 + \sqrt{H_0+36}$      \\
 \hline
0   &$\Delta$  &$y$    &$\phi$ &$24 + H_0 + 8\sqrt{H_0+36}$    \\
0 &$\Delta+1$&$y-2$  &$\pi$  &$H_0+96 + 16\sqrt{H_0+36}$  \\
0     &$\Delta+1$&$y$  &$\pi$   & $H_0+96+ 16\sqrt{H_0+36}$ \\
0     &$\Delta+1$&$y+2$  &$\pi$   & $H_0+96+ 16\sqrt{H_0+36}$  \\
0     &$\Delta+2$&$y$    &$S/\Sigma$   &$176 + H_0 + 24\sqrt{H_0+36}$   \\
 \hline
 \end{tabular}

\vskip .3cm

{\bf Table 5: Vector Multiplet I} \qquad $\Delta=\frac{5}{2} + \frac{1}{4}
\sqrt{H_0 + 36}$

\bigskip

 \begin{tabular}{|c|c|c|c|c|c|c|c|c|}\hline
&& &  spin & $\Delta^{(s)}$ &   $R$--symm.   &   field & Mass      \\
 \hline
 \hline
$\diamond$ &$\star$&& 1   &$\Delta+1$      &$y$
&$A/W$&$H_0+96 - 16\sqrt{H_0+36}$\\
 \hline
$\diamond$ &$\star$& $ \bullet$  & 1/2          &$\Delta+1/2$     &$y-1$    &$\lambda_{L}$         &$10 -  \sqrt{H_0+36}$  \\
$\diamond$ &$\star$&    &1/2               &$\Delta+1/2$     & $y+1$    & $\lambda_{L}$          & $10 -  \sqrt{H_0+36}$     \\
& $\star$ & &1/2          &$\Delta+3/2$     &$y-1$    &$\lambda_{T}$
&$6 - \sqrt{H_0+36}$  \\
&   &  &1/2&$\Delta+3/2$     &$y+1$    &$\lambda_{T}$
&$6 - \sqrt{H_0+36}$      \\\hline
$\diamond$ & $\star$& $\bullet$ & 0   &$\Delta$  &$y$    &$S/\Sigma$ &$176 + H_0 - 24\sqrt{H_0+36}$    \\
 &$\star$ & $\bullet$& 0 &$\Delta+1$&$y-2$  &$\pi$  &$H_0+96 -
16\sqrt{H_0+36}$  \\
$\diamond$ &  $\star$  & & 0     &$\Delta+1$&$y$  &$\pi$   & $H_0+96-
16\sqrt{H_0+36}$  \\
 &     & & 0     &$\Delta+1$&$y+2$  &$\pi$   & $H_0+96-
16\sqrt{H_0+36}$  \\
 &  & & 0     &$\Delta+2$&$y$    &$\phi$   &$24 + H_0 -8
\sqrt{H_0+36}$   \\\hline
 \end{tabular}

\vskip .3cm

{\bf Table 6: Vector Multiplet II} \qquad $\Delta=-\frac{3}{2} + \frac{1}{4}
\sqrt{H_0 + 36}$ 

\bigskip

\end{center}

\section{Classical $V_{(5,2)}$ cone equation and CFT}

Consider the non--compact four--fold defined by
\beq
\label{conifold}
\sum_{a=1}^5 z_a^2 = 0,
\eeq
which has an ordinary double point singularity at $z_a =0$ \cite{KW}.
This conifold is a cone over the homogeneous space $SO(5)/SO(3)$, that can
be retrieved by looking at the set of points at unit distance from the
singularity 
\beq
\sum_{a=1}^5 |z_a|^2 = 1.
\eeq
The full isometry group of this space is $SO(5) \times U_R(1)$
where the $U_R(1)$ plays the role of an $R$--symmetry group and
acts as a phase shift on the coordinates
\beq
\label{Qz}
z_a \to e^{i\a} z_a.
\eeq
Therefore the $z_a$ have  $Q = 1$ under this symmetry and transform
in the {\bf 5} of $SO(5)$.

Since it acts non--trivially on the canonical line bundle of the
conifold, the \eqn{Qz} transformation is an $R$--symmetry of the theory.
We can also see that it is an $R$--symmetry group from the fact
that the holomorphic 4--form
\beq
\Omega = \frac{{\rm d} z_1 \, {\rm d} z_2 \, {\rm d} z_3 \,{\rm d} z_4}{z_5}
\eeq
has $Q = 3$ under the $U_R(1)$ ($\Omega \to e^{3i\a}\Omega$).
The charge of the fermionic coordinates of
superspace is fixed by the requirement that they  transform as
$\sqrt{\Omega}$, and then $Q_{\theta} = \frac{3}{2}$.
Indeed, on a Calabi--Yau manifold we can always write the holomorphic form as
\beq
\Omega_{abcd} =  {}^t \eta \Gamma_{abcd} \eta,
\eeq
where $\eta$ is a covariantly constant spinor.
This means that $\Omega$ transforms as $\eta^2$, and
supersymmetries, being  generated by covariantly constant spinors,
transform as $\eta$.

As explained in sect. 3, it is convenient to fix the $R$--symmetry
value of the $\theta$ coordinates equal to one, and  to introduce the
rescaled $R$--charge $y = \frac{2}{3}Q$, under which $y_\theta =
1$.

In complete analogy with \cite{KW}, we can write a CY metric on
the cone by introducing the $SO(5)$ invariant K\"ahler potential
\beq
{\cal K} = \left(\sum_a \bar z_a z_a\right)^{3/4}.
\eeq
Defining  $r \equiv  \left(\sum_a \bar z_a z_a\right)^{3/8}$
and introducing a set of angular variables $y^A$, invariant under the
scaling of the $z$ coordinates, the metric can be put in the
standard form
\beq
d s_C^2 = d r^2 + r^2 g_{AB} d y^A d y^B \quad (A,B =
1,\ldots,7).
\eeq

This metric inserted in
\beq
ds_{11}^2 = r^2 \,(dx_0^2-dx_1^2-dx_2^2)+\frac{1}{r^2} \, ds_C^2
\eeq
plus the vacuum expectation value of the
three--form field strength ($F_4 \equiv dA_3$)
\beq
F_{\mu\nu\rho\s} = e \, \e_{\mu\nu\rho\s}
\eeq
describe the supergravity vacuum yielding the spontaneous compactification on a
seven--manifold from eleven to four space--time dimensions.

This supergravity solution has no moduli, as in eleven dimensional
supergravity there is no dilaton and the vev's of the fields giving the
$AdS_4 \times V_{(5,2)}$ compactification is uniquely fixed.
The only ``$\theta$--angle" we could introduce is a shift in the vacuum
value of the three--form $A_{\hat a\hat b\hat c}$ by a closed non--exact
three--form on the internal indices.
But we know that $H_3(V_{(5,2)},\IZ)$ is at most discrete torsion
\cite{Fetc}
and therefore there are no ``$\theta$--angles".
The absence of moduli reflects in the CFT definition implying that the
interacting fixed point is isolated in the parameter space.

This  seems to be  related to the geometrical nature of this
manifold.
It has been shown \cite{D} that, at variance with the $M^{111}$ and $Q^{111}$
cases, the Stiefel manifold does not admit a description in terms of toric
geometry and thus it is very difficult to see if it can be found as a partial
resolution of some orbifold.
If this could be done (like for the $Q^{111}$ manifold \cite{D}), it would
imply that there exists a flux from the orbifold CFT to this infrared point
\cite{KW,MP}, but it does not seem to be  the case.
Recent supergravity calculations \cite{AR} seem to confirm this fact at least
for fluxes connecting manifolds with the same topology.

\medskip

\nin
{\it The Conformal Field Theory}

\medskip

\nin
In the same spirit of \cite{KW}, the basic degrees of freedom of the desired
CFT can be understood upon ``solving" the \eqn{conifold} equation.
This can be done as follows: we set
\beq
\label{coord}
z^a={}^tA\ \Gamma^a B\equiv  {}^tA^{i}\ \Gamma^a_{ij} B^j
\eeq
where $A^i$ and $B^i$ are $SO(5)$ spinors (transforming in the fundamental
representation of $Sp(4)$) and $\Gamma^a$ are antisymmetric gamma matrices
in five dimensions, namely
\beq
\Gamma^a_{ij}=C_{ik} (\Gamma^a)^k{}_j ,
\eeq
$C_{ik}$ being the $Sp(4)$ invariant metric.
Since (using the identity $\Gamma^a_{ij} \Gamma_{a\,kl} = -6
C_{[ij}C_{k]l}+ C_{ij} C_{kl}$)
\beq
\sum_{a=1}^{5}z^a z_a\equiv ({}^tA\ \Gamma^a B)
({}^tA\ \Gamma_a B)\sim ({}^tA\ CB)^2,
\eeq
we have to supplement \eqn{coord} with the symplectic trace condition
\beq
\label{trace}
C_{ij}A^i B^j=0
\eeq
in order to retrieve the conifold equation \eqn{conifold}.

This matches exactly the representation of the conifold already used in
\cite{Fetc} in terms of the Pl\"ucker coordinates
\beq
\label{pluc}
p_{ij} = A_{[i} B_{j]},
\eeq
satisfying the Pfaffian constraint
\beq
\label{CCxx}
C^{[ij}C^{kl]} p_{ij} p_{kl} = 0,
\eeq
supplemented with the traceless condition $C^{ij} p_{ij} = 0$ \footnote{
Curiously \eqn{CCxx} is analogous to the moduli space of an $SU(2)$ $\cN = 2$
gauge theory with hypermultiplets with two flavours \cite{SWold}.}.
Equations \eqn{trace} and \eqn{pluc} are invariant under
$SL(2,\IC)$ transformations.
If we set $A^i=S^i_1$ and $B^i=S^i_2$, we see that the Pl\"ucker coordinates
\beq
p_{ij}=A_{[i} B_{j]}\equiv S^i_\alpha S^j_\beta
\epsilon^{\alpha\beta} \qquad\qquad \alpha,\beta=1,2
\eeq
and their symplectic trace $C^{ij}p_{ij}=0$ are invariant under
$SL(2,\IC)$.

Noting that $SL(2,\IC)$ is the complexification of $SU(2)$ \cite{Wess},
 we can gauge fix such invariance precisely by setting the
$SU(2)$ $D$--term to vanish
\beq
\label{Dterm}
D_{SU(2)}=0 \rightarrow \left\{ \bet{l}
$\sum_{i=1}^{4}|A^i|^2=\sum_{i=1}^4 |B^i|^2$, \\
\\
$\sum_{i=1}^4 A^i B^{*i}=0.$ \eet \right.
\eeq
The above discussion implies that the correct gauge group $\cG$ to be used
for $N$ coincident branes should reduce to $SU(2)$
for $N=1$.
Hence, choosing $\cG$ to reduce for $N=1$ to $SU(2)$, equation
\eqn{Dterm} fixes the $SL(2,\IC)$ residual invariance.
This gauge fixing is
quite analogous to the one used in \cite{KW}, where the complexification
of the $U(1)$ residual symmetry in the solution of the cone on $T^{11}$
is given by a complex rescaling of the relevant variables.

Given the above information, we can try to guess
 $\cG$,
when $N>1$. The product of two unitary group is excluded, since
the coordinate, as in $T^{11}$ (but not on the spheres), appears in the
KK spectrum and it is a gauge singlet so that the spinor
$S_i^\a$ must be in a pseudoreal
representation of $\cG$ (In a pseudoreal representation $C\bar S=S$, $C^2=-1$
and thus the gauge singlet is contained in the antisymmetric product
$(S\times S)_{asy}$).

We are thus led  as a minimal choice
to the product of the two non-simply laced groups
\beq
\cG= USp(2N)\times O(2N-1)
\eeq
The rationale for this choice is that, if we take the
singleton $S_i$ to be in the bifundamental representation of $\cG$, then,
since $S^i$ is in the $\bf 4$ of $Sp(4)$, the coordinate
\beq
\label{nacoord}
z^a=Tr\ (S^t \Gamma^a S)
\eeq
is non--zero only if the gauge group contains a factor $USp(2N)$;
the other factor must then be the orthogonal group $O(2N-1)$.
Indeed, the orthogonal group, having a symmetric invariant metric,
assures a non--zero value
for $z^a$ and moreover its order is fixed to be $2N-1$
because of the condition $\cG_{<vac>}=SU(2)$.

Such groups usually arise when one deals with orientifold projections
\cite{uranga} and for the present case $N$ refers correctly to the total number
of branes before mirroring.

From the chain decomposition
\bea
\label{catena}
USp(2N)\times O(2N-1)&\rightarrow& USp(2)\times USp(2(N-1))\times O(2(N-1))
\nonumber \\
&\rightarrow& USp(2)\times U(N-1)\times U(N-1)\\
&\rightarrow& USp(2)\times U(N-1)_{diag}\rightarrow USp(2)\times U(1)^{N-1}
\nonumber
\eea
we can retrieve the phase where all but one brane are free to move at smooth
points over the cone.
Looking at the chain \eqn{catena}, we see that, by the first
decomposition, we get
\beq
S^i=
\left( \bet{cc|ccc}  $A^i$ & $ B^i$ & 0 & \ldots & 0\\\hline
                  0  & 0   & &&\\
                  $\vdots$ & $\vdots$ & &$S^i_{A\Lambda}$& \\
                  0 & 0 & &&
            \eet       \right),
\eeq
where the upper left block is a $1\times  2$ matrix, the lower right block has indices
$A,\Lambda=1,\ldots, 2N-2$, while the off--diagonal blocks are rectangular
 $1\times
(2N-2)$ and $2 \times (2N-2)$ zero matrices.

Since $\;USp(2(N-1))$ and $O(2(N-1))$ both contain a $U(N-1)$
subgroup under which they both decompose as $(N-1)\oplus (\ol{N-1})$,
we have $A\rightarrow a,\bar a$ and $\Lambda\rightarrow \alpha,\bar\alpha$
($a,\bar a,\alpha,\bar\alpha=1,\ldots ,N-1$).
Correspondingly, the lower right $(2N - 2) \times (2N - 2)$ subblock of $S^i$
becomes \beq
S^i_{A\Lambda}=\left( \bet{c|c}
                        $S^i_{a\alpha}$ & $S^i_{a\bar\alpha}$\\\hline
                        $S^i_{\bar a \alpha}$ & $S^i_{\bar a \bar\alpha}$
              \eet     \right)
\eeq
which derives from  the second step of the chain \eqn{catena}.
Further going to the diagonal $U(N-1)$, we have
$S^i_{a\alpha}=S^i_{\bar a\bar\alpha}=0$ and setting $S^i_{a\bar\alpha}=A^i$,
$S^i_{\bar a\alpha}=B^i$ we find
\beq
\label{esse}
S^i=\left(\begin{tabular}{cc}
0&$A^i$\\
$B^i$&0\\
\end{tabular}\right).
\eeq

When we consider a generic vacuum configuration
$U_{diag}(N-1)\rightarrow U(1)^{N-1}$,  the $A^i$, $B^i$
subblocks reduce to diagonal (commuting) matrices in the Cartan subalgebra.

We remark that it is likely that there exist just {\sl one singleton}
$S^i$ and that $A^i$ and $B^i$ are just specific components of these $S^i$.
Indeed, promoting $A^i$ and $B^i$ to two independent singletons
$S^i,T^i$, would imply
that equation \eqn{nacoord}  admits the baryonic symmetry
\bea
\label{bario}
S^i&\rightarrow& S^i e^{i\alpha},\\
T^i&\rightarrow& T^i e^{-i\alpha}.
\eea
The baryonic symmetry
is related to the existence of $U(1)$ Betti multiplets in the KK spectrum
\cite{DF,CDD,Fetc},
which only occur if there are non-trivial Betti numbers $b_i,\ i\neq 1,7$.
However,  $V_{(5,2)}$ has the same real homology  the seven-sphere
$S^7$,  and thus a continuous baryonic symmetry is ruled out.

Thus we propose  that the CFT describing  a large number of
$M2$--branes on the \eqn{conifold} singularity is given by
the infrared fixed point of an $USp(2N)\times O(2N-1)$
gauge theory where the basic degrees of freedom  are chiral multiplets
 $S^i$  lying in
the {\bf 4} of $SO(5)$, with $R$--symmetry charge $Q = 1/2$ (or $y =
1/3$) and in the $(2N\,,\,2N-1)$ irrep.  of the gauge group.
In the brane construction of gauge theories usually there can be other
matter fields in symmetric and antisymmetric representations.
We assume here that such representations decouple at the conformal
IR fixed point.

The chiral fields (singletons) of the conformal field theory
 have $\Delta = |y| = \frac{1}{3}$.
This means that flowing to the interacting point they
acquire an anomalous dimension $\gamma = - \frac{1}{6}$.
This makes the conformal dimension violate the unitarity bound
$\Delta \geq \dfrac{1}{2}$, but since the singleton field is not
a gauge
group singlet it is not an observable of the theory.
The analogous phenomenon occurs for the five--dimensional case
$T^{11}$, where $\Delta_{A,B} = \frac{3}{4} < 1$ and for the proposed
CFT's dual to the seven--dimensional manifolds $M^{111}$ and $Q^{111}$
\cite{Fetc}.

As already remarked, the gauge theory exists only in the ultraviolet limit
where it is not conformal and where
the gauge vector potential, which is a singlet of the matter group $SO(5)$
is in the adjoint representation of $USp(2N) \times O(2N-1)$.
We could  dualize it, at least in the Coulomb branch, and
then reintroduce it in the CFT.
However, from the KK analysis, we see that we have no states
corresponding to products of this true singleton field (with
$\Delta = \frac{1}{2}$) and therefore we have no coordinates for
the Coulomb branch.

As we will see later, it is also essential to introduce a
superpotential whose Jacobian ideal gives the needed vanishing
relations for the correct matching of the chiral primaries with the
supergravity hypermultiplets.
This is  given generically by the sixth power of the singleton fields
\beq
\label{superpot}
{\cal W}(S_i) = C_{ijklmn} \, {\rm Tr} (S^i S^j S^k S^l S^m S^n).
\eeq
where the tensor $C_{ijklmn}$ is constructed by an appropriate linear
combination of products of three $Sp(4)$ invariant metrics $C_{ij}$.
It should probably be made of a combination of the following
structures
\bea
\label{suno}
& Tr\left[ (SS)\ (SS)\ (SS) \right]& \\
\label{sdue}
& Tr\left[ (SS)\ S \Gamma^a S\ S\Gamma_a S\right] & \\
\label{stre}
& Tr\left[  S\Gamma^a S\ S\Gamma^{bc} S\ S\Gamma^{de} S\right]\ 
\epsilon_{abcdef} & \\ 
\label{squattro}
&Tr\left[ (SS)\ S\Gamma^{ab} S\ S\Gamma_{ab} S \right]& \\
\label{scinque}
&Tr\left[ S \Gamma^{ab} S\ S\Gamma_{bc}S\ S\Gamma^c_a S \right]& \\
\label{ssei}
&Tr\left[ S\Gamma^a S\ S\Gamma^b S\ S\Gamma_{ab}S\right].
\eea

Let us consider the previous structures for $N = 1$, when we
can drop the trace symbol.
We easily see that they are the six a priori existing singlets which
can be obtained from the product of six $\bf 4$ spinor representations
of $Sp(4)$.
Next we note that when \eqn{esse} holds, all the
above structures are given by products of three $A^i$ and three $B^i$
contracted with three $C_{ij}$ tensors.
Actually, there is just one possible
$Sp(4)$ invariant that can be built, namely
\beq
(A^i B^j C_{ij})^3
\eeq
Furthermore, the first four structures
\eqn{suno}--\eqn{squattro} are antisymmetric under the exchange of
$A^i$ and $B^i$ while the last two are symmetric. That means that, by Fierz
identities, \eqn{suno}--\eqn{squattro} must be proportional to each other,
while the other two must vanish identically.

\bigskip

\section{AdS/CFT correspondence}

\nin
{\it $OSp(4|2)$ conformal superfields}

\medskip

\nin
A generic $OSp(4|2)$ representation \cite{CFN} is labelled by three quantum
numbers, according to the $OSp(4|2) \sim SO(2) \times SO(3) \times
U_R(1)$ decomposition of the supergroup.  
They are the energy $\Delta$, the spin $s$ and the $R$--charge $y$.

This generic representation is unitary if 
\beq
\label{ubound}
\Delta \geq 1+s+|y|,
\eeq
while short chiral representations can occur for
\beq
\Delta =|y| \geq \frac{1}{2}.
\eeq

Like in the $SU(2,2|1)$ case \cite{CDDF}, at the threshold of the unitarity
bound  \eqn{ubound}, we can obtain short representations.
These BPS--saturated states correspond to short superfields which
thus satisfy some differential constraint.

Operators with protected dimensions are related to such shortenings
and they fall in three categories:
\begin{itemize}

\item {\bf Chiral superfields}: They occur when $\Delta = |y|$ and satisfy
the condition 
\beq
\ol{D}_\a \Phi(x,\theta,\bth) = 0,
\eeq
or ${D}_\a \Phi(x,\theta,\bth) = 0$ for anti--chiral ones.

\item {\bf Conserved currents}: They occur when $\Delta = 1+s$ and satisfy
\bea
&& D^{\alpha_1} J_{\alpha_1 \ldots \alpha_{2s}}(x,\theta,\bth) = 
\ol{D}^{\alpha_1} J_{\alpha_1 \ldots \alpha_{2s}}(x,\theta,\bth) = 0  \qquad
{\rm if}\ s \neq 0\\
&{\rm or}& D^2 J(x,\theta,\bth) = \ol{D}^2 J(x,\theta,\bth) =
0. \qquad {\rm for}\ s = 0.
\eea

\item {\bf Semiconserved currents}: They occur when $\Delta = 1+s+|y|$ and
satisfy 
\bea
\label{semi}
&& \ol{D}^{\alpha_1} L_{\alpha_1 \ldots \alpha_{2s}}(x,\theta,\bth) =  0,
\qquad (s \neq 0)\\ && \ol{D}^2 L(x,\theta,\bth) = 0, \qquad (s = 0)
\eea
if left--semiconserved, or the conjugate conditions if
right--semiconserved.

\end{itemize}
It is trivial to see that a right and left semi--conserved
superfield is also conserved.

\medskip

\nin
{\it The protected operators}

\medskip

\nin
From the CFT point of view, we expect to have chiral operators corresponding to
the wave--functions of the conifold \cite{KW}.
Such operators are given by
\beq
\label{Phik}
{\rm Tr}\, \phi^k \equiv Tr \left(z^{a_1} \ldots z^{a_k}\right)\, C_{a_1
\ldots a_k}
\eeq
with $C_{a_1\ldots a_k}$  a completely symmetric and traceless rank $k$ tensor.
They have $\Delta = y = \frac{2}{3}k$.

Surely, there should be a conserved current related to the
global $SO(5)$ symmetry,  which should be a singlet of the gauge
and $R$--symmetry group and that we can identify as
\beq
J^{ab} \equiv \bar S \Gamma^{ab} S
\eeq
This $J^{ab}$ should be massless and satisfy $\bar D^2 J^{ab}=
D^2J^{ab} = 0$.
Its conformal dimension is therefore $\Delta = 1$.

Another operator with protected dimension we certainly expect is
given by the stress--energy tensor
\beq
J_{\a\b} = \bar{D}_\a \bar S D_\b S +\bar{D}_\b \bar S D_\a S +
i \bar S
\stackrel{\leftrightarrow}{\partial\!\!\!\slash}{}_{\a\b} S,
\eeq
which has $\Delta = 2$, $y = 0$ and satisfies $D^\a J_{\a\b} =
\bar D^\a J_{\a\b} = 0$.

It is now trivial to see that we should also expect KK supergravity
states corresponding to the following semi--conserved superfields
\beq
\label{graviton}
{\rm Tr}\, (J_{\a\b} \phi^k )\qquad {\rm and} \qquad {\rm Tr}\,(J^{ab} \phi^k)
\eeq
(or the conjugate ones).

It seems more problematic to find the appropriate singleton
combinations which appear as semiconserved spin $1/2$ superfields in
the CFT corresponding to short gravitino multiplets on the
supergravity side.
In the theory at hand there is no field like the $W_\a$ of the
$T^{11}$ case \cite{KW,CDDF} and thus there is no natural candidate for these
operators.  We also have to be careful not to use simple descendants
of primary operators and this makes the task more difficult.
Anyway, once we have the isometry group quantum numbers,
from the KK analysis, we can see that the appropriate combinations
of the $S^\alpha_i$ are uniquely fixed, and will be written explicitely below.

\medskip

\nin
{\it The correspondence}

\medskip
\nin
Given the structure of the $OSp(4|2)$ multiplets of
eleven--dimensional supergravity compactified on $AdS_4 \times V_{(5,2)}$, we
can make the comparison between these results and the CFT predictions.
We can also make use of these results to explicitly determine the expression of
the fermionic operators related to the short gravitino multiplets.

Along the lines of the five--dimensional case of type IIB supergravity on 
$AdS_5 \times T^{11}$ \cite{CDDF,CDD}, we look for rational conformal
dimensions occurring in the KK multiplets and see whether they correspond to
the right shortenings needed to be related to the previously described
conformal operators.

From the energy values of the multiplets, it is easy to see that a rational
conformal dimension can be obtained only if $H_0 +36$ or $H_0+24$ are squares
of  rational numbers.

As in the $T^{11}$ case, we obtain rationality when we saturate the bound
on the $R$--charge of a given harmonic, i.e. when in the Young Tableaux all
the boxes which can be charged have the same $R$--charge.
This occurs for the representations\footnote{In the form [Young
indices]$_{charge} = [M+N,M]_y$.} $[k \; ,\; 0]_{\frac{2}{3}k}$
of $SO(5)_{U_R(1)}$ in the $H_0 + 36$ case and for
$[k \; ,\; 0]_{1+\frac{2}{3}k}$ in the $H_0 + 24$ case.
The corresponding square roots are given by $ 6+\frac{8}{3}k$ and
$4+\frac{8}{3}k$ respectively.
We have solved the rationality constraint for the more generic case of
$[m+n+k \; , \; m]_{1+\frac{2}{3}k}$, and
we have found that there are two other infinite series of operators with
rational dimension, for $m$ and $n$ satisfying the following relations
\bea
m^2 - n^2 -2 mn -3n -m =0,&& {\rm for\ }H_0 + 36, \\
m^2 - n^2 +2m(1-n) =0, && {\rm for\ } H_0 +24.
\eea
This gives sequences of numbers with no simple rationale.
Anyway we will see that as for $T^{11}$, beside the case $m=n=0$,
only another couple of $SO(5)\times U_R(1)$ quantum numbers, is
related to shortenings, while all the others correspond to the rational long
multiplets partially noticed in \cite{G} and completely clarified in
\cite{CDDF}.
Here these couples are $m=1,n=0$ and $m=1,n=1$ respectively.

Let us now introduce these conditions on the $SO(5) \times U_R(1)$ quantum
numbers in the $\Delta $ values of the supergravity multiplets and see when the
shortening occurs.

We start with the graviton and vector multiplets for which we have some
expectations to be verified and then pass to the gravitino multiplets.
The graviton multiplet has
\beq
\Delta = \frac{1}{2}+\frac{1}{4}\sqrt{H_0 + 36}.
\eeq
If the $SO(5) \times U_R(1)$ irrep is $[k \; , \;0]_{\frac{2}{3}k}$,
it reduces to
\beq
\Delta = 2 + \frac{2}{3}k,
\eeq
which is the shortening condition $\Delta = 1+s+|y|$
related to the protected operator
\eqn{graviton} corresponding to the massless and short graviton multiplets.

It can be easily seen that also in the $[k+1 \; , \;1]_{\frac{2}{3}k}$ case,
it is obtained a rational state with $\Delta = 3 + \frac{2}{3}k$.
These states do not satisfy the shortening condition $\Delta = 1+s+|y|$, but
they can be put in correspondence with the rational non supersymmetry protected
operators\footnote{Here and in the following the conformal operators have to be
understood as projected along the $SO(5)$ Young tableaux of the corresponding
KK state.}
\beq
{\rm Tr}\,(J_{\a\b}J^{ab} \phi^k).
\eeq

For the vector II,
\beq
\Delta = -\frac{3}{2}+\frac{1}{4}\sqrt{H_0 + 36}.
\eeq
If we choose the $[k \; ,\; 0]_{\pm\frac{2}{3}k}$ irrep., we obtain states with
\beq
\Delta = \frac{2}{3}k,
\eeq
which are hypermultiplet ($\Delta = |y|$) states associated to the $\phi^k$ operators.
When the $G$--irrep is $[k+1 \; ,\; 1]_{\frac{2}{3}k}$, we obtain again a shortening of
the multiplet.
Its anomalous dimension is given by
\beq
\Delta = 1 + \frac{2}{3}k,
\eeq
and is related to the massless gauge vector multiplet of the $SO(5)$ matter
group or to short vector multiplets corresponding to Tr$(J^{ab} \phi^k)$ operators.

The other type of vector multiplets we have found never undergo shortening, but
we can easily find the CFT rational long operators.
Their anomalous dimension is
\beq
\Delta = \frac{5}{2}+\frac{1}{4}\sqrt{H_0 + 36},
\eeq
which for the $[k \; ,\; 0]_{\pm\frac{2}{3}k}$ irreps reduces to
\beq
\Delta = 4 + \frac{2}{3}k
\eeq
and for the $[k+1 \; ,\; 1]_{\pm\frac{2}{3}k}$ case reduces to
\beq
\Delta = 5 + \frac{2}{3}k.
\eeq

It is easy to see that the related CFT operators are given by
\beq
{\rm Tr}\,(J_{\a\b}J^{\a\b}\phi^k)
\eeq
and
\beq
{\rm Tr}\,(J_{\a\b}J^{\a\b}J^{ab}\phi^k).
\eeq

Let us now examine the shortening conditions for the gravitino multiplets.
Type I gravitino has
\beq
\Delta = -\frac{1}{2}+\frac{1}{4}\sqrt{H_0 + 24},
\eeq
which, for the $[k \; ,\; 0]_{1+\frac{2}{3}k}$ irreps reduces to
\beq
\Delta = \frac{1}{2} + \frac{2}{3}k=-\frac{1}{2}+|y|.
\eeq
This does not correspond to a shortening condition, but nevertheless satisfies
the unitarity bound $\Delta \geq 1+s+|y|$.

We can obtain unitary multiplets when the $G$ quantum numbers are $[k+2\; ,\;
1]_{1+\frac{2}{3}k}$.
In this case indeed
\beq
\label{Lgen}
\Delta = \frac{5}{2} + \frac{2}{3}k=1+s+|y|
\eeq
and therefore we obtain short gravitino multiplets.

For type II gravitino we have
\beq
\Delta = \frac{3}{2}+\frac{1}{4}\sqrt{H_0 + 24},
\eeq
which, for the $[k \; ,\;0]_{1+\frac{2}{3}k}$ irreps reduces to
\beq
\label{Xgen}
\Delta = \frac{5}{2} + \frac{2}{3}k
\eeq
undergoing shortening, and for $[k+2 ,\; 1]_{1+\frac{2}{3}k}$
\beq
\Delta = \frac{9}{2} + \frac{2}{3}k
\eeq
gives long rational multiplets.

Having the $OSp(4|2)$ and matter group quantum numbers, we can try to guess
the corresponding conformal operators. For $k = 0$, those related to the short
type I gravitinos are given by \beq
\label{elle}
{\rm Tr}\,L_\a = Tr\left[ \left( \bar S \G^a S \,
\bar D_\a \bar S \G^{bc} S -
\bar D_\a \bar S \G^a S \, \bar S \G^{bc} S \right)
\right],
\eeq
while those related to short type II gravitinos are
\beq
\label{ics}
{\rm Tr}\,X_\a = Tr\left[ S \G_a S \,  \left( \bar S \G_b \bar S \,
\bar D_\a  \bar S \G^{ab} S - 2\bar S \G_b \bar D_\a  \bar S \,
 \bar S \G^{ab} S
 \right)\right],
\eeq
which become Tr$(L_\a \phi^k)$ and Tr$(X_\a \phi^k)$ for the generic
cases \eqn{Lgen} and
\eqn{Xgen} respectively.
Equations \eqn{elle} and \eqn{ics} are easily seen to obey the
semi--conservation condition \eqn{semi}.

We point out that in the $L_{\a}$ operator, only the irreducible
$a[bc]$ representation survives once we use the $\G$--matrices
identities and the $D$--term equations.

Let us note explicitly that, as anticipated in the introduction the
type II  short gravitino multiplet $Tr\ X_\alpha$ has a lowest component
of $R$--symmetry $y=1$, so that its $\bar\theta$ component, which we call
$\tilde W_\alpha$, has $y=0$. Moreover, it is a singlet under
$SO(5)$ so that in the infrared limit $\tilde W_\alpha$ has the same quantum
numbers (apart from the conformal dimension) as the original gauge field in
the ultraviolet limit.

We may compare this result with the gravitino sector superfield of the
four dimensional SCFT dual to the $T^{11}$ compactification of Type IIB theory
and called $L^{1k}_{\dot\alpha}$ in \cite{CDDF}. There, the vector field
strength superfield $W_\alpha$ is the singleton of the conformal theory,
so that it does not appear in the spectrum of $T^{11}$ compactification.
In the present case instead, $Tr\ [X_\alpha \phi^k]$ does indeed appear in the
spectrum of $V_{(5,2)}$ being a composite field of singletons. 
Furthermore, $L^{1k}_{\dot\alpha}$ does not exist for $k=0$, since in this
case it would reduce to \cite{CDDF} 
\beq
L^{10}_{\dot\alpha}=Tr\ (e^V W_{\dot\alpha} e^{-V})
\eeq
which vanishes identically while $Tr\ [X_\alpha \phi^k]$ is different
from zero even for $k=0$.

Finally, for type II gravitinos,
we have in addition  states with $\Delta = \frac{9}{2} + \frac{2}{3}k$
corresponding to long rational multiplets, which can be written as
\beq
{\rm Tr}\,\left[(S\bar D_\b \bar S)(S\bar D^\b \bar S) \bar L_\a \phi^k\right].
\eeq

\section{Summary}

In order to collect our results, we present a table where we list the
multiplet type as well as spin, representation and energy of the highest
states for  $M$--theory compactified on the Stiefel manifold, and match them
with the boundary conformal superfields. 
These results merely rely on the AdS/CFT correspondence.

It remains an open problem to make an explicit construction of the
ultraviolet description of the underlying field theory in terms of $D2$--brane
gauge theory. 

\begin{center}

\bet{||c|c|c|c|c||}\hline\hline
s & $\Delta$ & \bet{c} $SO(5)_{U_R(1)}$ \\ irreps\eet & multiplet &
\bet{c} Conformal \\ superfield \eet\\\hline\hline
1 & $2+ \dfrac{2}{3} k$ & $[k\, ,\, 0]_{\frac{2}{3}k}$ & \bet{c} short \\
graviton$^\star$ \eet & $T_{\a\b}
\phi^k$ \\\hline
1 & $3+ \dfrac{2}{3} k$ & $[k+1\, ,\, 1]_{\frac{2}{3}k}$ & \bet{c} long \\
graviton \eet & $J^{ab} \, T_{\a\b}\phi^k$ \\\hline\hline
1/2 & $\dfrac{1}{2} + \dfrac{2}{3} k$ & $[k\, ,\, 0]_{1+\frac{2}{3}k}$ &
\bet{c} non \\ unitary \eet &  \\\hline
1/2 & $\dfrac{5}{2}+ \dfrac{2}{3} k$ & $[k\, ,\, 0]_{1+\frac{2}{3}k}$ &
\bet{c} short \\ gravitino II \eet & $X_\a \phi^k$ \\\hline
1/2 & $\dfrac{5}{2}+ \dfrac{2}{3} k$ & $[k+2\, ,\, 1]_{1+\frac{2}{3}k}$ &
\bet{c} short \\ gravitino I\eet & $L_\a\phi^k$ \\\hline
1/2 & $\dfrac{9}{2}+ \dfrac{2}{3} k$ & $[k+2\, ,\, 1]_{1+\frac{2}{3}k}$ &
\bet{c} long \\gravitino I\eet & $(S \bar D_\b \bar S) (S \bar D^\b \bar S)
\bar L_\a\phi^k$ \\\hline\hline
0 & $\dfrac{2}{3} k$ & $[k\, ,\, 0]_{\frac{2}{3}k}$ & hypermultiplet & $\phi^k$ \\\hline
0 & $1+ \dfrac{2}{3} k$ & $[k+1\, ,\, 1]_{\frac{2}{3}k}$ & \bet{c} short \\
vector$^\star$ II\eet & $J^{ab}\, \phi^k$ \\\hline
0 & $4+ \dfrac{2}{3} k$ & $[k\, ,\, 0]_{\frac{2}{3}k}$ & \bet{c} long \\
vector I\eet & $T_{\a\b} T^{\a\b}\,\phi^k$ \\\hline
0 & $5+ \dfrac{2}{3} k$ & $[k+1\, ,\, 1]_{\frac{2}{3}k}$ & \bet{c} long \\
vector I\eet & $T_{\a\b}T^{\a\b}\, J^{ab}\,\phi^k$ \\\hline\hline
\eet

$^\star$ Massless for $k=0$.
\end{center}


\vskip 1truecm

\paragraph{Acknowledgements.}
We are glad to thank  M. Bianchi, L. Castellani, D. Fabbri, P. Fr\'e,
A. Sagnotti and especially Y. Oz and  A. Uranga for interesting
discussions. S. F. is grateful to the Politecnico of Torino
for the kind hospitality during various stages of this work. S.F. is
supported by the DOE under grant DE-FG03-91ER40662, Task C.
This work is also supported by the European Commission TMR programme
ERBFMRX-CT96-0045 (University and Politecnico of Torino and Frascati nodes).

\section*{Appendix A: Rescaled connection and curvature on G/H }
\setcounter{equation}{0}
\makeatletter
\@addtoreset{equation}{section}
\makeatother
\renewcommand{\theequation}{A.\arabic{equation}}

In this appendix we  present an algebraic technique to derive the rescaled
connection and curvature on a coset manifold given the structure constants of
the $G$, $H$ and $G/H$ groups\footnote{The results of this section were
derived in collaboration with L.Castellani.} that generalizes the formulae of
\cite{CRW2}.
The $a,b$ are the coset indices, $i,j$ are the $H$ indices while
$e^a$ and $\omega^i$ are the vielbeins.

The Maurer--Cartan equations for $e^a$ and $\omega^i$ are
\begin{subequations}
\label{MC}
\bea
d e^a + \dfrac{1}{2} {C^a}_{bc} e^b e^c +  {C^a}_{bi}  e^b
\omega^i &=& 0,\\
d \omega^i + \dfrac{1}{2} {C^i}_{bc} e^b e^c +  \frac{1}{2} {C^i}_{jk}
\omega^j \omega^k &=& 0.
\eea
\end{subequations}

Under a rescaling of $e^a$, equations \eqn{MC} become:
\begin{subequations}
\label{MCrescaled}
\bea
d e^a + \dfrac{1}{2} \frac{r(b)r(c)}{r(a)} {C^a}_{bc} e^b e^c +
\frac{r(b)}{r(a)} {C^a}_{bi}  e^b
\omega^i &=& 0,\\
d \omega^i + \dfrac{1}{2} r(a)r(b) \, {C^i}_{bc} e^b e^c +  \frac{1}{2} {C^i}_{jk}
\omega^j \omega^k &=& 0.
\eea
\end{subequations}

The connection one--form on $G/H$ can be defined by
\beq
\label{defcon}
d e^a - {\cB^a}_b e^b = 0.
\eeq

Combining \eqn{defcon} and \eqn{MCrescaled} yields
\beq
\cB^{a}{}_b =  \frac{1}{2} \frac{bc}{a} {C^a}_{bc} e^c + \frac{r(b)}{r(a)}
\,  {C^a}_{bi} \omega^i + {K^a}_{bc} e^c,
\eeq
where ${K^a}_{bc}$, symmetric in $b,c$, is determined by the
requirement of antisymmetry of $\cB$.

Thus the antisymmetric connection $\cB$ is given by
\beq
\cB^{a}{}_b =  \frac{1}{2}  {\IC^a}_{bc} e^c  \frac{r(b)}{r(a)}
\,  {C^a}_{bi} \omega^i
\eeq
where
\beq
\label{IC}
{\IC^a}_{bc} \equiv \frac{r(b)r(c)}{r(a)} {C^a}_{bc} +
\frac{r(a)r(c)}{r(b)} {C^f}_{ce} \eta^{ea} \eta_{fb} -
\frac{r(a)r(b)}{r(c)} {C^g}_{fb} \eta^{af} \eta_{cg}.
\eeq

The Riemann curvature is defined in terms of the connection as
\beq
{R^a}_b \equiv d {\cB^a}_b - {\cB^a}_c {\cB^c}_b \equiv {R^a}_{bde}
e^d e^e.
\eeq

Substituting the definition of $\cB$ in terms of the structure
constants given above, and using the Maurer--Cartan equations for the
differentiated vielbeins and Jacobi identities for products of
structure constants, one arrives at
\bea
\label{Riemann}
{R^a}_{bde} &=& - \frac{1}{4} {\IC^a}_{bc} {C^c}_{de} \frac{r(d)
r(e)}{r(c)} - \frac{1}{2}
{C^a}_{bi}{C^i}_{de} r(d)r(e) + \nonumber \\
&-& \frac{1}{8} {\IC^a}_{cd} {\IC^c}_{be}
+ \frac{1}{8} {\IC^a}_{ce} {\IC^c}_{bd}.
\eea

This form of the Riemann tensor is more general than the one
presented in \cite{CRW2}, where the final result depended only on
the ${C^a}_{bc}$ and not on the ${\IC^a}_{bc}$ due to the hypothesis
that the Killing metric be completely diagonal.
In our case instead the mixed components $\gamma_{ia}$ are non--zero, while
the condition that within an isotropy--irreducible subspace the Killing metric
is proportional to $\delta_{ab}$ still holds.
This is necessary to ensure the antisymmetry of the connection $\cB^{ab}$.

It is straightforward to verify that when the Killing metric is diagonal, the
${\IC^a}_{bc}$ reduces to the combination $\left(\!\!\bet{c} $a\ b$ \\ $c$
\eet\!\!\right) {C^a}_{bc}$ of \cite{CRW2}.

\section*{Appendix B:The Reduction of $SO(7)$ under $SO(3) \times SO(2)$.}
\setcounter{equation}{0}
\makeatletter
\@addtoreset{equation}{section}
\makeatother
\renewcommand{\theequation}{B.\arabic{equation}}

In this section we reduce the $SO(7)$ indices to $H$--irreducible
indices.

The embedding of  $SO(3) \times SO(2)$ in $SO(7)$ is defined by
\beq
(T_H)^{ab} = ({C_H}^{\a})^{\b} (T_{\a\b})^{ab},
\eeq
relating the generators of $H$ in an $SO(7)$ irrep to the generators
of $SO(7)$ in the same irrep through the structure constants.
In the vector representation of $SO(7)$ one has
$$
(T_{\a\b})^{\g\delta} = - \delta_{\a\b}^{\g\delta}
$$
and therefore
$$
(T_H)^{\a\b} = ({C_H}^{\a})^{\b}.
$$

Using the expressions for the structure constants
one obtains
\beq
(N)_{\a}{}^{\b} = \left( \bet{c|c|c}
 0 & ${\delta_m}^{\hm}$ & 0 \\\hline
 $-{\delta_{\hm}}^{m}$ & 0 & 0 \\\hline
 0 & 0 & 0
  \eet\right),
\eeq
and
\beq
(J^i)^{\a\b} = \left( \bet{c|c|c}
 $\e^{imn}$ & 0 & 0 \\\hline
 0 & $\e^{i\hm\hn}$ & 0 \\\hline
 0 & 0 & 0
  \eet\right).
\eeq

Thus the $SO(7)$ vector reduces under $H$ as
\beq
{\bf 7} \to {\bf 3}_1 \oplus {\bf 3}_{-1} \oplus {\bf 1}_0,
\eeq
where the first number labels the $SO(3)$ irrep, while the second one
is the $U(1)_H$ charge.

To construct them in the spinor representations we use the following
$\g$ matrices:
\bes
\g^{m} &=& \left\{ i \s^{1} \otimes \s^1 \otimes \s^2, i \s^1
\otimes \s^2 \otimes \unity , -i \s^1 \otimes \s^3 \otimes \s^2
\right\}, \\
\g^{\hm} &=& \left\{ -i \s^{2} \otimes \s^2 \otimes \s^1, i \s^2
\otimes \s^2 \otimes \s^3 , -i \s^2 \otimes \unity \otimes \s^2
\right\}, \\
\g^7 &=& i\s^3 \otimes \unity \otimes \unity.
\ees
The charge conjugation matrix is
\beq
C = \s^1 \otimes \unity \otimes \unity.
\eeq

The $N$ generator in the spinor rep. is thus
\beq
N = \frac{1}{2} \g^{m} \g^{\hm} \delta_{m\hm} = - \frac{i}{2} \left( \bet{cccc|cccc}
1 &&&&&&& \\
 & 1 &&&&&& \\
 && 1 &&&&& \\
 &&& - 3 &&&&\\\hline
 &&&& -1 &&& \\
 &&&&& -1 && \\
 &&&&&& -1 & \\
 &&&&&&& 3\eet\right),
\eeq
and the $J^{i}$ are
\beq
J^{i} \sim \left( \bet{c|c}
\bet{c|c} $\e^{ijk} \phantom{\dfrac{C}{C}}$ & 0 \\\hline 0 & 0 \eet & {\msbm O}$_4$ \\\hline
{\msbm O}$_4$  & \bet{c|c} $\e^{ijk}\phantom{\dfrac{C}{C}}$ & 0 \\\hline 0 & 0 \eet
\eet\right),
\eeq
so the eight--dimensional spinor representation of $SO(7)$ reduces
under the $H$ subgroup as
\beq
{\bf 8} \to  {\bf 3}_{1/2} \oplus {\bf 3}_{-1/2} \oplus {\bf 1}_{3/2}.
\eeq

We will decompose the eight--component Majorana spinor
as $\left( \bet{c} $H$ \\ $\Lambda$
\eet \right)$, where
$$\Lambda = \left( \bet{c} $\phi_{(3, 1/2)}$ \\ $\omega_{(1, -3/2)}$
\eet\right).
$$
Since ours are Majorana spinors $C \Xi^{*} = \Xi$, $H^* = \Lambda$
and then our generic spinor is
\beq
\Xi = \left( \bet{c} $\phi_-^{k}$ \\ $\omega_+$ \\  $\phi_+^{k}$ \\ $\omega_-$
\eet \right).
\eeq


%
%

\end{document}